\documentclass[aps,prc,showpacs,twocolumn,superscriptaddress]{revtex4-1}
\usepackage{amsmath,amsfonts,amssymb,amsthm,float,mathastext, array, booktabs, tabularx, mathtools}
\usepackage{booktabs}
\usepackage{graphicx}
\graphicspath{{fig/}}
\usepackage{siunitx}
\usepackage{hyperref}
\hypersetup{hidelinks,colorlinks=false,breaklinks=true,urlcolor=ocre}
\usepackage{dcolumn}% Align table columns on decimal point
\usepackage{bm}% bold math
\usepackage[english]{babel}
\usepackage[autostyle]{csquotes}
\usepackage{color}
\begin{document}
\preprint{APS/123-QED}

\title{Exploring Charge Transport Dynamics in a Cryogenic P-Type Germanium Detector}% Force line breaks with \\
\author{P. Acharya}
\affiliation{Department of Physics, The University of South Dakota, Vermillion, SD 57069, USA}%
\author{M. Fritts}
\affiliation{School of Physics and Astronomy, University of Minnesota, Minneapolis, MN, 55455, USA}
\author{D.-M. Mei} 
 \email{Corresponding author.\\Email: Dongming.Mei@usd.edu}
 \affiliation{Department of Physics, The University of South Dakota, Vermillion, SD 57069, USA}%
%\author{V. Mandic}
%\affiliation{School of Physics and Astronomy, University of Minnesota, Minneapolis, MN, 55455, USA}
\author{G.-J. Wang}
\affiliation{Department of Physics, The University of South Dakota, Vermillion, SD 57069, USA}
\author{R. Mahapatra}
\affiliation{Department of Physics $\&$ Astronomy, Texas A $\&$ M University, College Station, TX 77843, USA}
\author{M. Platt}
\affiliation{Department of Physics $\&$ Astronomy, Texas A $\&$ M University, College Station, TX 77843, USA}
\date{\today}% It is always \today, today,%  but any date may be explicitly specified
\begin{abstract}

This study explores the dynamics of charge transport within a cryogenic P-type Ge particle detector, fabricated from a crystal cultivated at the University of South Dakota (USD). By subjecting the detector to cryogenic temperatures and an Am-241 source, we observe evolving charge dynamics and the emergence of cluster dipole states, leading to the impact ionization process at 40 mK. Our analysis focuses on crucial parameters: the zero-field cross-section of cluster dipole states and the binding energy of these states. For the Ge detector in our investigation, the zero-field cross-section of cluster dipole states is determined to be $8.45 \times 10^{-11}\pm 4.22\times 10^{-12}~cm^2$. Examination of the binding energy associated with cluster dipole states, formed by charge trapping onto dipole states during the freeze-out process, reveals a value of $0.034 \pm 0.0017$ meV. These findings shed light on the intricate charge states influenced by the interplay of temperature and electric field, with potential implications for the sensitivity in detecting low-mass dark matter.

\end{abstract}
%\pacs{29.40.Mc, 24.10.-i, 29.85.Fj, 13.75.-n}% PACS, the Physics and Astronomy
\keywords{Suggested keywords}%Use show keys class option if keyword
\maketitle
%\tableofcontent
\section{Introduction}
Recent astronomical investigations consistently highlight the pivotal role of dark matter (DM) in the cosmos, characterized by its non-luminous and non-baryonic nature, constituting the majority of the universe's material composition~\cite{kahlhoefer2017review,porter2011dark,brink2005beyond,agnes2018low}. Strong evidence supports the notion that DM influences the cosmos by creating expansive halos around galaxies, as observed in the Milky Way~\cite{agertz2016impact,frampton2009did,diemand2007dark,iocco2015evidence}. The quest for direct evidence lies within laboratory research, exploring potential interactions between DM particles and ordinary matter beyond gravitational forces. While weakly interacting massive particles (WIMPs) have been a focus, recent theoretical frameworks like asymmetric DM and dark sectors prompt exploration of low-mass DM particles. Thus, the MeV-scale particle emerges as a noteworthy candidate for low-mass DM~\cite{mei2018direct,angloher2017results,an2018directly}, yet detecting them poses challenges. The need for detectors with exceptional sensitivity to discern a solitary electron-hole (e-h) pair complicates matters due to the narrow energy span involved. Overcoming this hurdle requires leveraging modern technology and innovative methodologies for precise and reliable capture and quantification of these minute energy signals. The relentless pursuit of sophisticated detectors and pioneering detection techniques continues as researchers delve deeper into the mysteries surrounding low-mass DM~\cite{barbeau2022report}. These advancements hold the promise of unraveling one of the universe's profound enigmas, potentially bringing us closer to understanding its intricate tapestry.

Owing to their exceptional sensitivity, germanium (Ge) detectors have emerged as a promising avenue for addressing the challenge of low-mass DM detection, presenting a compelling alternative to conventional methodologies ~\cite{gascon2021low}. These detectors possess a distinct advantage for probing low-mass DM due to their remarkable ability to efficiently generate electron-hole pairs, each requiring an average energy of 3 eV. Complementing this characteristic is Ge's narrow band gap of $\sim$0.7 eV, further bolstering its suitability for such investigations at millikelvin (mK) temperatures ~\cite{oh2021cryogen}. Leveraging the concept of doping, it becomes possible to significantly expand the parameter space for detecting low-mass DM using Ge detectors. By judiciously introducing impurities into the Ge matrix, particularly shallow-level impurities boasting binding energies around 0.01 eV, a fascinating phenomenon emerges: the creation of dipole states and cluster dipole states in conditions below 6.5 K ~\cite{mei2022evidence}.
What distinguishes these dipole states and cluster dipole states is their binding energy, which plunges even lower than that of the impurities themselves ~\cite{raut2023development,bhattarai2023investigating}. This intriguing aspect opens a pathway to potentially detecting low-mass DM. 

Yet, despite well-explored knowledge concerning the binding energies of impurities in Ge, a conspicuous gap remains regarding the binding energies of these dipole states and cluster dipole states at cryogenic temperatures, especially those hovering below 100 mK. When Ge is cooled to these frigid temperatures, as experienced in the vicinity of liquid helium, a compelling process unfolds: the expulsion of residual impurities from its conduction or valence bands. These expelled impurities then find solace in localized states, giving rise to electric dipoles (denoted as $D^{0^*}$ for donors and $A^{0^*}$ for acceptors) or neutral states ($D^0$ and $A^0$). These states signify excited neutral impurity configurations, tethered by binding energies less than 10 meV. The captivating implication here is the capacity of these dipole states to ensnare charge, thereby orchestrating the formation of cluster dipole states ($D^+$ and $D^-$ for donors, and $A^+$ and $A^-$ for acceptors), which exhibit even lower binding energy levels. The specifics of these states depend on the operational temperature ~\cite{mei2022evidence,acharya2023observation,klinkigt2013cluster}.

\begin{figure} [htbp]
  \centering
  \includegraphics[clip,width=0.9\linewidth]{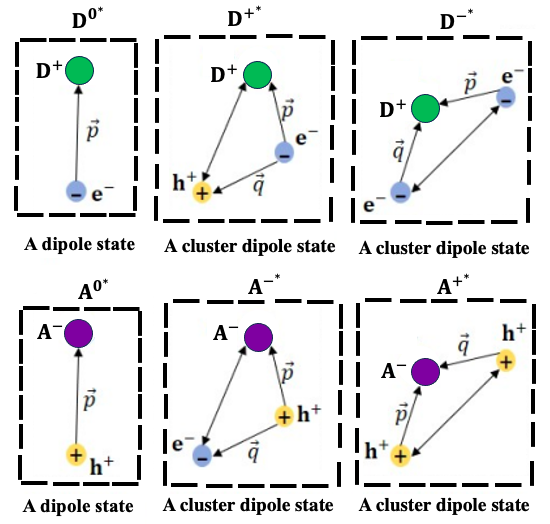}
  \caption{The procedural framework delineating the progression towards the genesis of both excited dipole states and cluster dipole states within a dual-tiered Ge detector — encompassing an upper tier (n-type) and a lower tier (p-type) — operational within sub-10 K temperatures, is embodied by the subsequent stages. Herein, $\vec{p}$ and $\vec{q}$ symbolize the distinct dipole moments affiliated with each state.}
  \label{fig:f0}
\end{figure}

As temperatures decrease, a remarkable exponential reduction in the density of free charge carriers becomes apparent~\cite{mei2023exploring}. Approaching temperatures lower than 10 K, donor (or acceptor) atoms predominantly maintain their unionized configuration, confining fifth electrons (or vacant holes) within these atomic entities. The extent of this encapsulation is governed by the Onsager radius, denoted as $R=\frac{e^2}{4\pi\varepsilon\varepsilon_{0} K_{B}T}$, where $\varepsilon$ represents the relative permittivity of Ge (with a value of 16.2), $\varepsilon_{0}$ is the permittivity of free space, $K_{B}$ is the Boltzmann constant, and $T$ is the temperature. Remarkably, at temperatures below 10 K, $R$ can significantly exceed the dimensions of the donor or acceptor atom~\cite{sze2021physics}.
Consequently, the fifth electrons (or vacant holes) associated with the donor (or acceptor) atoms may undergo thermal dissociation from the atomic nucleus, initiating the formation of electric dipoles due to the segregation of opposing charges~\cite{mei2023exploring}. This sequential progression leads to the generation of dipole states with the capacity to trap charge through Coulombic attraction, giving rise to the emergence of cluster dipole states~\cite{mei2022evidence}. As depicted in Figure \ref{fig:f0}, this visualization illustrates the formation of excited dipole states and cluster dipole states in n-type Ge or p-type Ge, respectively.
Within an excited dipole state, whether associated with a positively charged donor ion or a negatively charged acceptor ion, a state of profound confinement prevails due to the lattice deformation potential. Consequently, the phase space available for capturing charge carriers is inherently more constrained than that of bound electrons (or holes), which retain the capability to traverse within the bounds set by the Onsager radius (R)~\cite{mei2022evidence, mei2023exploring}.

This intricate interplay foretells a higher likelihood of generating $D^{+^{*}}$ states (or $A^{-^{*}}$ states) within an n-type detector (or a p-type detector) than the likelihood of yielding $D^{-^{*}}$ states (or $A^{+^{*}}$ states)~\cite{mei2023exploring}. This asymmetry implies that within an n-type detector, holes undergo more pronounced entrapment effects compared to electrons, whereas electrons experience greater susceptibility to entrapment than holes within a p-type detector ~\cite{arnquist2022alpha,meijer2003solution}. At the University of South Dakota (USD), our research has delved extensively into the binding energy characteristics of cluster dipole states, as documented in recent studies~\cite{bhattarai2023investigating,raut2023development}. This investigation unfolded at a low temperature of 5.2 K, uncovering a binding energy threshold consistently below 10 meV, a parameter intricately linked to the prevailing electric fields. Notably, our analysis has spotlighted the pivotal role played by the Onsager radius in shaping the spatial confines conducive to the emergence of cluster dipole states, with temperature exerting a significant influence on this parameter. The intricate interplay between temperature and the Onsager radius serves as a governing factor in determining the binding energy of cluster dipole states. Our research transcends the microscopic realm, carrying macroscopic implications. It reveals that lower temperatures engender an expansion of the Onsager region, consequently leading to a reduction in binding energies. Conversely, higher temperatures induce a contraction of this region, resulting in heightened binding energies. Our commitment to advancing this knowledge extends further, as we seek to validate these findings across distinct temperature regimes, including conditions exceeding 800 mK and those as low as 40 mK. Through this empirical exploration, we aim to offer concrete evidence for the intricate thermal dynamics that influence the binding energy within cluster dipole states.

 \section{Experimental Methods and Observed Physical Phenomenon}

\begin{figure} [htbp]
  \centering
  \includegraphics[clip,width=0.9\linewidth]{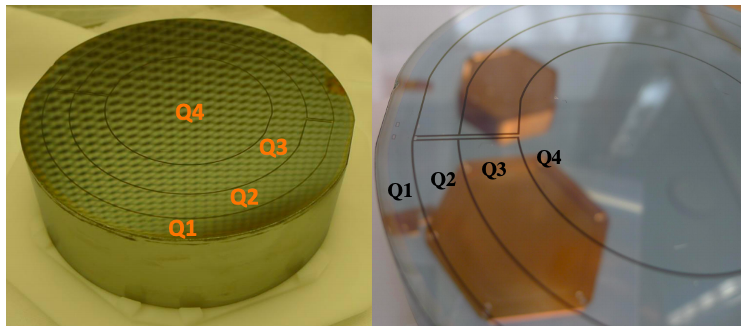}
  \caption{This illustration features the detector scrutinized in our investigation, equipped with four read-out electrodes labeled as $Q_{1}$, $Q_{2}$, $Q_3$, and $Q_4$. The detector boasts dimensions of 10 cm in diameter and 3.3 cm in thickness, endowing it with a mass of approximately 1.34 kg.}
  \label{fig:f1}
\end{figure}

Leveraging cutting-edge infrastructure, USD boasts an avant-garde platform for crystal growth and detector development. Central to this system is the employment of a zone refining technique, artfully harnessed to attain exceptional purification levels in commercial ingots ~\cite{wang2015crystal,yang2015zone}. These meticulously purified ingots lay the foundation for crystal growth through the renowned Czochralski method. With over a decade of research and development experience, we have achieved increased consistency in growing detector-grade crystals. It is noteworthy that we have integrated advanced machine learning techniques into this process to further enhance the growth of large-size detector-grade crystals, aiming to attain a level of purity conducive to exploring low-mass DM phenomena~\cite{acharya2023machine}.
The crux of this study lies in a crystal cultivated in 2014 at USD, revealing a net impurity level of $|N_A-N_D|=4\times10^{11}~cm^{-3}$. This particular crystal assumes a p-type configuration. The subsequent fabrication of the detector, orchestrated at Texas A $\&$ M University, boasts a configuration encompassing four channels for charge readout positioned atop the detector. Departing from conventional paradigms, the detector's grounding component integrates a uniform aluminum (Al) electrode as opposed to a grid arrangement. This innovative design choice, while safeguarding the electric field's integrity, acts as a bulwark against charge leakage, thereby sustaining an unwavering and uniformly distributed electric field. Notably, this departure might have implications for the detector's susceptibility to effective neutralization through light-emitting diodes (LEDs).
The detector's geometric layout mirrors the essence of SuperCDMS-style detectors, as pictured in Figure \ref{fig:f1}. Following meticulous wire-bonding, the detector seamlessly integrates into a dilution refrigerator, embarking on a comprehensive testing regimen hosted at the K100 Detector Testing Facility at the University of Minnesota (UMN).

\begin{figure} [htbp]
  \centering
  \includegraphics[clip,width=0.9\linewidth]{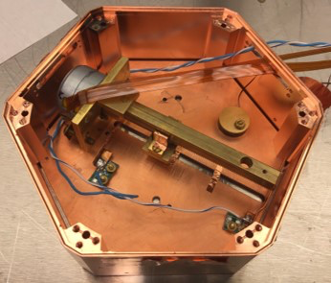}
  \caption{Presented here is the arrangement of the $^{241}$Am source manipulator, ingeniously crafted utilizing superconducting mobile coil technology. This groundbreaking design facilitates meticulous control and positioning of the radioactive source, thereby significantly augmenting its effectiveness across a spectrum of scenarios ~\cite{mast2020situ}.}
  \label{fig:f3}
\end{figure}

The phenomenon of impact ionization prompted by impurities within Ge specimens has received extensive attention from researchers, particularly at temperatures surpassing 4 K ~\cite{zylbersztejn1962theory,pickin1978impact}. Noteworthy contributions to this knowledge pool have been made by various scholars. Recently, Phipps et al. have achieved a significant breakthrough by investigating the impact ionization of impurities using SuperCDMS-style detectors at an ultracold temperature of $\sim$40 mK ~\cite{phipps2016observation}. Furthermore, F. Ponce et al. demonstrated a room temperature pulsed laser's effect on a SuperCDMS silicon HVeV detector, unveiling coherent probabilities for charge trapping and impact ionization within the high-purity Si substrate under specific conditions ~\cite{ponce2020measuring}. Aligned with these pioneering works, our endeavor delves into the realm of time-dependent impact ionization, mirroring the ultracold environment at $\sim$40 mK. We employ a detector constructed from a distinctive and ultra-stable Ge crystal. This detector experiences gradual cryogenic cooling, eventually reaching an extraordinary temperature of approximately 40 mK. The ensuing meticulous evaluations transpire over two separate refrigeration runs: Run 67 in 2018 and Run 74 in 2021. In the context of Run 67, the arrangement involved situating four $^{241}$Am sources directly above individual channels on the detector, an arrangement aptly illustrated in Figure~\ref{fig:f1}. These setups employ lead collimators sporting 0.2 mm apertures, strategically permitting the transmission of 59.54 keV $\gamma$ rays while adeptly blocking alpha particles through source encapsulation. Our findings manifest in the spectra from each channel, displaying distinct 59.54 keV peaks, with comprehensive measurements conducted diligently over a two-week interval. Contrastingly, Run 74 deploys a singular $^{241}$Am source, ingeniously mounted on a carriage facilitated by a superconducting stepper motor, as graphically represented in Figure \ref{fig:f3}. This inventive source design incorporates a 0.5 mm collimator hole, leading to a reduced flux of incident $\gamma$ rays on the detector, constituting approximately 75\% of the sources utilized in Run 67. Notably, the two runs unfold over extended time spans, inherently introducing fluctuations in the detector's operational state. These fluctuations can wield considerable influence over the detector's charge collection efficiency and trapping characteristics. For an all-encompassing comprehension of the nuanced intricacies underpinning these experiments, an in-depth analysis is available in the research paper authored by Acharya et al ~\cite{acharya2023observation}.

\section{Experimental Data Analysis and Results}

 \begin{figure}[htbp]
  \centering
  \includegraphics[clip,width=0.9\linewidth]{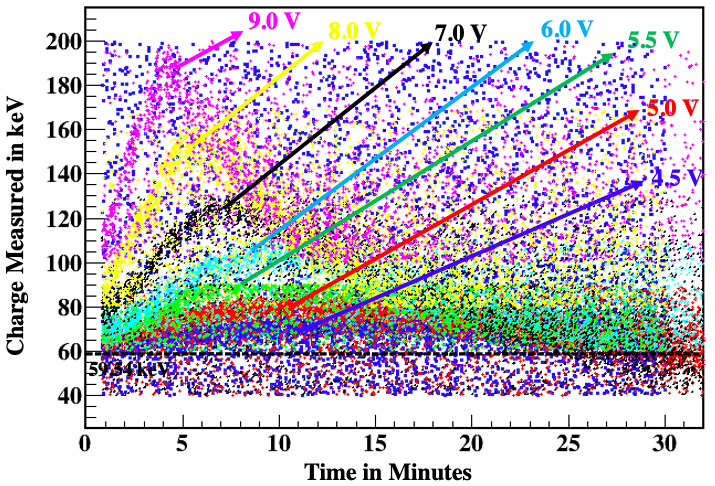}
  \caption{Depicted here is the evolving charge response from Q4, discernible across time when subjected to favorable biases during Run-67. The foundational signal associated with the 59.54 keV gamma-ray emission from Q4 exhibited an initial linear rise within a matter of minutes, which swiftly transitioned into an exponential decline spanning tens of minutes.}
  \label{fig:f4}
\end{figure}

 \begin{figure}[htbp]
  \centering
  \includegraphics[clip,width=0.9\linewidth]{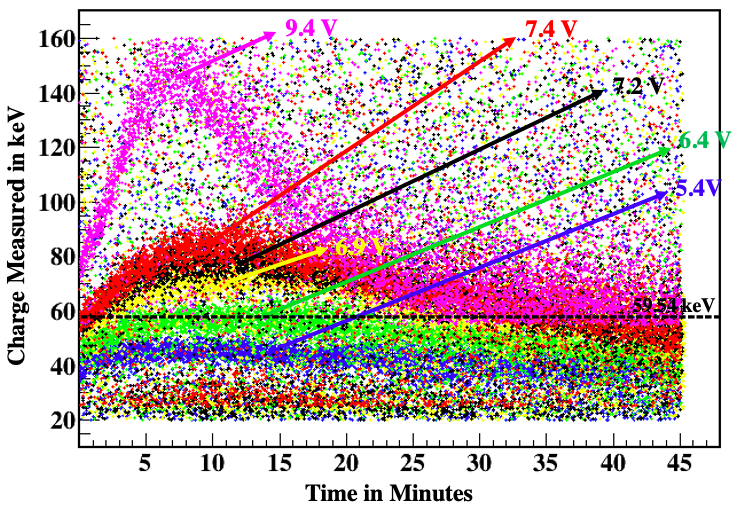}
  \caption{Provided is the temporal progression of the charge response in Q4, witnessed under positive biases throughout Run-74. The initial climb of the 59.54 keV gamma ray baseline stemming from Q4 displayed a linear trajectory during the initial moments, rapidly transitioning into an almost exponential decay pattern spanning a maximum duration of approximately 10 minutes.}
\label{fig:f5}
\end{figure}

We embarked on a thorough analysis, utilizing data collected at various bias voltages to comprehensively delineate the charge collection efficiency for each channel, scrutinizing its interplay with bias and time. This investigation has yielded intriguing insights into the behavior of the 59.54 keV calibration line, as depicted in Figure~\ref{fig:f4} and Figure~\ref{fig:f5}. Our observations unveiled a multitude of time-dependent phenomena: signal amplitudes from the 59.54 keV line exhibiting an initial increase post-bias application, succeeded by a subsequent descent towards a steady-state value. This distinctive phenomenon exclusively manifested under positive biases, owing to the strategic positioning of $^{241}$Am sources on the biased side of the crystal, instigating the immediate collection of electrons and the subsequent drift of holes across the entirety of the detector's thickness.

Remarkably, this effect displayed varying degrees of potency across channels, with its most pronounced manifestation occurring in the central channel (Q4), while its presence was faintest in the outer ring channel (Q1), indicative of a radial dependence. In light of this, our study centered its focus on the 59.54 keV line within the central channel (Q4) under positive biases, as it showcased the most promising outcomes.

Upon delving into the equilibrium state at t=0, where $p-n=N_{D}^{+}+N_{A}^{-}=0$, representing the charge neutralization state ~\cite{hagstrum1961theory,wei1994ion}, with p signifying free holes, $n$ indicating free electrons, $N_{D}^{+}$ representing positively ionized donor atoms, and $N_{A}^{-}$ representing negatively ionized acceptor atoms. We also introduce expressions for the density of acceptor impurities: $p+p_d=N_A$, where $p_d$ signifies the unionized holes bound to the atom; similarly, $n+n_d=N_D$, with $n_d$ representing the unionized donor electrons. When weaving these expressions together, they coalesce into a coherent depiction.

Figure~\ref{fig:f4} and Figure~\ref{fig:f5} visually articulate the temporal evolution of the impact ionization phenomenon, underscored by the drift of hole carriers under positive bias conditions. The process unfolds with hole impact ionization exhibiting a linear surge, modulated by bias and event rate. This amplified charge signal experiences an exponential decay, the timescale of which is intrinsically linked to the prevailing bias. This behavior is attributed to the attenuation of charge collection efficiency, stemming from the disruption of the bulk electric field due to the presence of trapped charges. The linear enhancement in the signal arising from hole impact ionization, in conjunction with the absence of comparable electron impact ionization, suggests that the spectrum of plausible sites for hole impact ionization is modest at its inception, originating from the capture of drifting electrons.

This intricate phenomenon encompasses the interplay of three distinct processes: $e^{-} + A^{0^{*}} \rightarrow A^{-^{*}}$, $h^{+} + A^{-^{*}} \rightarrow e^{-} + 2h^{+} + A^{-}$, and $h^{+} + A^{-} \rightarrow A^{0^{*}}$. The first process entails the creation of cluster dipole states as a consequence of electron migration within the detector, spurred by background radiation. The second corresponds to the ionization of cluster dipole states initiated by hole carriers, while the third involves the confinement of $h^{+}$ carriers, culminating in the cessation of neutralization in the series' denouement ~\cite{acharya2022observation}. These intricate mechanisms find graphical representation in Figure~\ref{fig:f0}.

\section{The Theoretical Framework}
To comprehensively understand the intricate nuances influencing the detector's performance under low-temperature conditions, it is crucial to subject various theoretical models to rigorous testing. The insights gained from these systematic evaluations have the potential to provide invaluable information, shaping our understanding of the detector's efficacy in low-mass DM detection. We have extensively explored several theoretical frameworks, supplementing our investigation with calculations. A detailed presentation of these comprehensive findings and discussions are outlined below.

\subsection{Impact Ionization}

At low cryogenic temperatures, the Ge detector showcases intriguing phenomena related to impact ionization (Fig.~\ref{fig:f4}). 
Our earlier analysis of the impact ionization phenomenon employed a well-defined physical model that accounted for the observed behavior of charge carriers. We have established a relationship between the detected charge energy, denoted as $E(t)$, and the input 59.54~keV $\gamma$ rays through the following equation ~\cite{acharya2023observation}:
 \begin{equation}
     \label{e1}
     E(t) = E_{\gamma} \{p_{0}+ p_{1}exp[\frac{p_{2}}{p_{3}}(1-exp(-p_{3} t))]\}exp(-p_{4}t),
 \end{equation}
where $E_{\gamma}$ = 59.54 keV. Our focus lies on two parameters, $p_{0}$ and $\&$ $p_{1}$, where $p_{0} + p_{1}$ signifies the impact ionization factor, specifically for absolute impact ionization observation. If $p_{0} + p_{1} >1$, it indicates a gain in charge energy due to impact ionization.

The sole instance of observed absolute impact ionization transpired at $t=0$ during Run-67. Consequently, our analysis will center on understanding the detector's distinctive signature for impact ionization at low temperatures, leveraging the data obtained from Run-67. Upon applying the data from Run-67 to fit the model outlined in Equation~\ref{e1}, the fitted parameters for $p_{0}$ 
  and $p_{1}$
  under various biases were determined as,

\begin{table}[htbp]
\centering
\begin{tabular}{|p{1.5cm}|p{1.5cm}|p{1.8cm}|}
\hline
\multicolumn{3}{|c|}{Run 67} \\
\hline
\multicolumn{1}{|c|}{Bias} & \multicolumn{1}{c|}{$p_{0}$} & \multicolumn{1}{c|}{$p_{1}$} \\
\hline
4.5 V & 1.05 & 0.015 \\
5 V & 1.07 & 0.018 \\
5.5 V & 1.09 & 0.019 \\
6 V & 1.13 & 0.020 \\
7 V & 1.22 & 0.028 \\
8 V & 1.28 & 0.032 \\
9 V & 1.39 & 0.038 \\
\hline
\end{tabular}
\caption{Summary of the fitting for parameters $p_{0}$ and $p_{1}$ under various biases in Run-67 ~\cite{acharya2023observation}.}
\label{t1}
\end{table}

 \begin{figure} [htp]
  \centering
  \includegraphics[clip,width=0.95\linewidth]{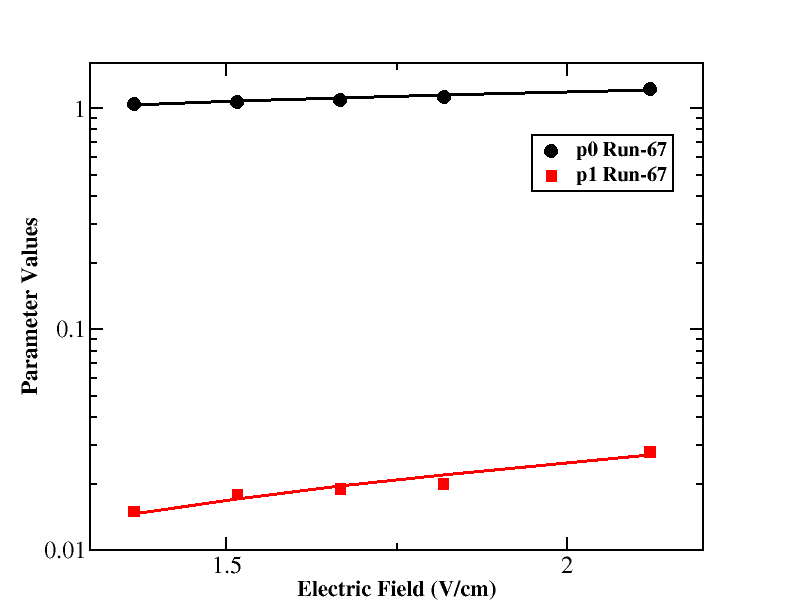}
  \caption{Shown are the fitting parameters, $p_0$ and $p_1$ using equation \ref{e1} as a function of the applied field, $E$, and fits a linear regression model, which demonstrates the correlation of the parameters with the electric field using the data from Run-67. The fitting functions for those parameters in Run~67 are: $p_{0}$=0.25$E$+0.69, $p_{1}$=0.017$E$-0.0087.}
  \label{f6}
\end{figure}
  
Figure ~\ref{f6} represents the correlation between two parameters, $p_0$ and $p_1$, and the applied electric field at 40 mK. Regression analysis demonstrates a strong correlation, indicating these parameters vary according to the applied bias. The sum of these parameters, $p_0+p_1$, defines an amplification factor, $A$. If $n_t$ represents the charge carriers at any time t, it can be expressed as $n_t=n_0e^{t/\tau_{imp}}$, where $n_0$ denotes the charge carriers at time $t=0$ ~\cite{mei2023exploring}. Here, $\tau_{imp}=\frac{1}{N_{a}\sigma_{imp}(E)v_d}$ stands for the charge generation time through impact ionization. Here, $N_a$ represents the impurity concentration, $\sigma_{imp}(E)$ is the impact ionization scattering cross-section as a function of the electric field, and $v_d$ signifies the drift velocity, given by $v_d=d/t$, where $d$ denotes the detector width and $t$ is the drift time. It should be noted that the amplification factor, $A= n_t/n_0$. After incorporating the equations $n_t=n_0e^{t/\tau_{imp}}$ and $\tau_{imp}=\frac{1}{N_{a}\sigma_{imp}(E)v_d}$ alongside the amplification factor, the derived equation elucidating impact ionization due to the cluster dipole at cryogenic temperatures is succinctly expressed as:

\begin{equation}\label{len}
log(A)=N_a\times\sigma_{imp}(E)\times d,
\end{equation}

\subsection{Zero Field Cross-Section}

The zero-field impact ionization cross-section for charged carriers within a detector denotes the detector's effective interaction area in the absence of an external electric field and quantifies the likelihood per unit area of charged carriers inducing an impact ionization event within the detector material. This parameter is intricately impacted by several factors, including the characteristics of the particle (such as charge and energy), alongside the specific properties inherent to the detecting material. Notably, the zero-field cross-section for cluster dipoles unveils a captivating domain within low-temperature physics. Understanding and harnessing this cross-section hold considerable potential for advancing our comprehension of elusive particles and their interaction dynamics within the context of low-temperature environments. If $\sigma_0$ denotes the zero field cross-section, its mathematical expression can be written as ~\cite{palmier1972impurity,gutierrez2000low,phipps2016ionization,sundqvist2012carrier}:

\begin{equation}\label{cross}
\sigma_0= \frac{\pi}{2}\times\frac{\hbar^{2}}{2m_h^*E_{BB'}},
\end{equation}

Here, $m_h^*$ represents the effective mass of holes in the conduction band, and $E_{BB'}$ denotes the binding energy for the cluster dipole state, which will be discussed in detail later in the section dedicated to calculating the binding energy of cluster dipole states, as defined by Equation ~\ref{bind_clu}. Considering $m_h^* = 0.21m_e$, where $m_e$ represents the mass of an electron, equivalent to $0.51~MeV/C^2$. We can plug these parameters into Equation \ref{cross} to obtain $\sigma_0 = \frac{2.85 \times 10^{-15}}{E_{BB'}}~cm^2$, where $E_{BB'}$ has the unit in eV.

\subsection{Impurity Freeze-Out}

In a recent study, Mei et al. demonstrated a freeze-out process in both an n-type detector with N${d}$ = 7.02$\times$10$^{10}$/cm$^3$ and a p-type detector with N${a}$ = 6.2$\times$10$^{9}$/cm$^3$~\cite{mei2022evidence, mei2023exploring}. In these detectors, as the temperature falls below 11 K, there is a notable decrease in relative capacitance in the Ge detectors. Simultaneously, impurities in the detectors freeze out of the conduction or valence band, leading to a decrease in free charges within the detector volume as it is cooled down to 6.5 K~\cite{mei2023exploring}. Further cooling towards 5.2 K maintains a constant capacitance, indicating the absence of free charges. Nearly all impurity atoms then form electric dipole states ($D^{0^{*}}$ for donors and $A^{0^{*}}$ for acceptors)~\cite{mei2023exploring}.

In the case of a p-type Ge detector, if an impurity atom is in its ground state, it is unable to capture charges. Nevertheless, when the detector is cooled to 5.2 K, this impurity atom undergoes a transition to a dipole state, gaining the ability to capture charges and forming a cluster dipole state.  During this freeze-out process, impurity atoms, previously responsible for generating free carriers above 11 K, undergo a transition into bound states. These bound states manifest as electrically neutral, reflecting the transformation of impurities in the Ge detector into localized states. This phenomenon is commonly referred to as "freeze-out" in the context of semiconductor physics~\cite{mei2023exploring}.

\begin{figure}[H]
\centering
\includegraphics[clip,width=0.9\linewidth]{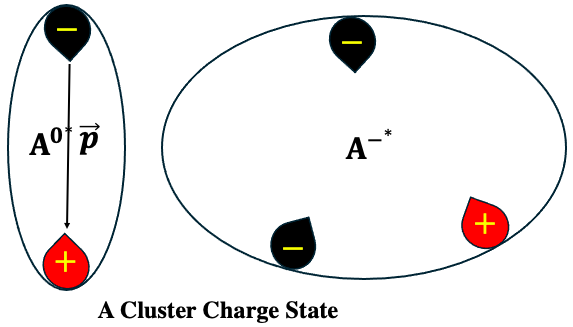}
\caption{Shown are a p-type dipole state and a p-type cluster dipole state.}
\label{cluster}
\end{figure}

If the same detectors continue to cool down, as the detector temperature reaches 40 mK, it is anticipated that the dipole moments of these dipole states will increase, influenced by the temperature-dependent size of dipoles. These enhanced dipole states may weakly trap charge carriers, potentially resulting in the formation of cluster dipole states ($D^{+^{*}}$, $D^{-^{*}}$ for donors, and $A^{+^{*}}$, $A^{-^{*}}$ for acceptors)~\cite{mei2022evidence, raut2023development, bhattarai2023investigating}.
Figure ~\ref{cluster} illustrates both a p-type dipole state and a p-type cluster dipole state~\cite{mei2023exploring}.

\begin{figure}[H]
\centering
\includegraphics[clip,width=0.9\linewidth]{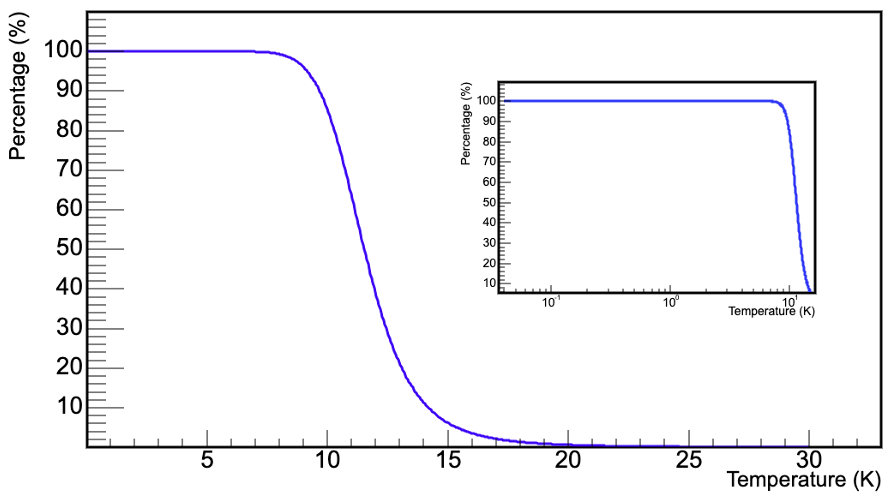}
\caption{The presented graph illustrates the temperature-dependent ratio of holes bound to the acceptors, with $N_A=4\times 10^{11}~cm^{-3}$ and $E_a - E_v \simeq 0.01~eV$ as the parameters for generating this plot. Additionally, it's important to observe that the inset plot conveys the same information across a wider temperature range, displayed in a logarithmic scale.}
\label{frac}
\end{figure}

The freeze-out phenomenon significantly influences the behavior of Ge detectors at extremely low temperatures. In the state of freeze-out, the proportion of holes associated with acceptor levels in a p-type material with an impurity density of $N_a$ is ~\cite{mei2023exploring}:
\begin{equation}\label{h_frac}
\frac{p_a}{p+p_a}=\frac{1}{\frac{N_V}{4N_a}exp[-\frac{E_a - E_v}{K_BT}]+1},
\end{equation}
where $p$ signifies the number density of free holes in the valence band and $p_a$ denotes the number density of holes bound to acceptors, the mathematical relationship is expressed as $N_a = p + p_a$. Here, $N_V$ signifies the density of states in the valence band, and $E_a - E_v \simeq 0.01~eV$ represents the ionization energy of acceptors. The fraction of holes bound to acceptors, as defined by Equation ~\ref{h_frac}, is visually represented in Figure ~\ref{frac}. 

The graph vividly illustrates a significant rise in the percentage of holes bound to acceptors within the temperature range of 20 K to 8 K. It's important to note that this freeze-out range in temperature slightly deviates from the observations in Mei et al.\cite{mei2022evidence, mei2023exploring} due to variations in detector impurity levels. This dependence on impurity levels is also evident from Equation\ref{h_frac}. By rearranging the terms, Equation ~\ref{h_frac} can be reformulated as:
\begin{equation}\label{p_hole}
p=\frac{p_aN_V}{4N_a}exp[-\frac{E_a - E_v}{K_BT}],
\end{equation}

\begin{figure}[H]
\centering
\includegraphics[clip,width=0.9\linewidth]{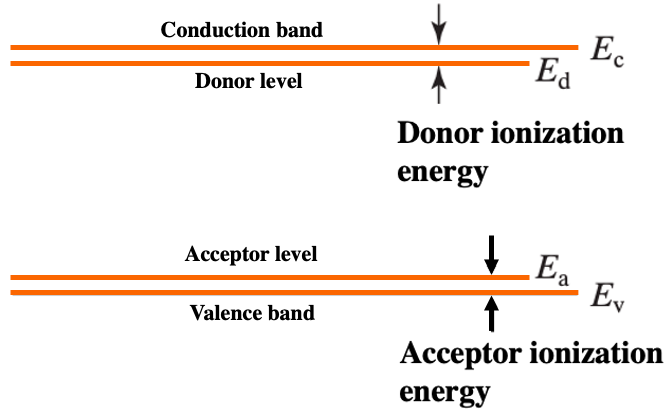}
\caption{Displayed are the energy levels of donors and acceptors within semiconductor materials ~\cite{ferry2013semiconductors,berger2020semiconductor}.}
\label{band}
\end{figure}

While residual impurity atoms remain ionized until approximately 20 K, a distinctive phenomenon emerges within the temperature range of 20 K to 8 K. During this period, these impurity atoms undergo a freeze-out process, transitioning into bound states that manifest as effectively charge-neutral entities. This abrupt transformation is intricately linked to the band structures illustrated in Figure \ref{band}, where $E_c - E_d \simeq 0.01$ eV and $E_a - E_v \simeq 0.01~eV$ ~\cite{mei2020impact}.

The abrupt transition depicted in Figure~\ref{frac} highlights the intricate relationship between temperature fluctuations and the corresponding band structures. It reveals a significant shift in the behavior of impurity atoms within this particular temperature range (20 K to 8 K).
 Under such circumstances, the expression for the number density of free holes ($p$) existing within the valence band can also be delineated as ~\cite{mei2018direct,van1950theory}:
\begin{equation}\label{pp_hole}
p=[\frac{N_aN_V}{4}]^{1/2}exp[-\frac{E_a - E_v}{2K_BT}],
\end{equation}
where $N_V$ is the effective density of states in the valence band for Ge materials, which is given by $N_V=9.6\times10^{14}T^{3/2}~cm^{-3}$.

In situations where the temperature drops below 8 K for the p-type detector under investigation, the number density of free holes is significantly smaller, by several orders of magnitude, than the number density of holes bound to acceptors. Consequently, it is reasonable to assume that $p_{a}$ equals $N_{a}$ and $E_a - E_v = E_B$, where $E_B$ is the binding energy of the bound states. Therefore, combining Equations ~\ref{p_hole} and \ref{pp_hole} facilitates the determination of the binding energy for the bound states - dipole states:
\begin{equation}\label{bin}
E_{BB}=K_BT~log(\frac{N_V}{4p_a}),
\end{equation}

Please note that the binding energy of dipole states, represented by $E_{BB}$, differs from the ionization energy of acceptors, as illustrated in Figure~\ref{band}. This distinction arises due to the formation of dipole states within a particular range of the Onsager radius when holes bound to acceptors. Consequently, the binding energy of these dipole states, influenced by the size of the dipole, may differ from the ionization energy of impurity atoms.

\begin{figure}[H]
\centering
\includegraphics[clip,width=0.9\linewidth]{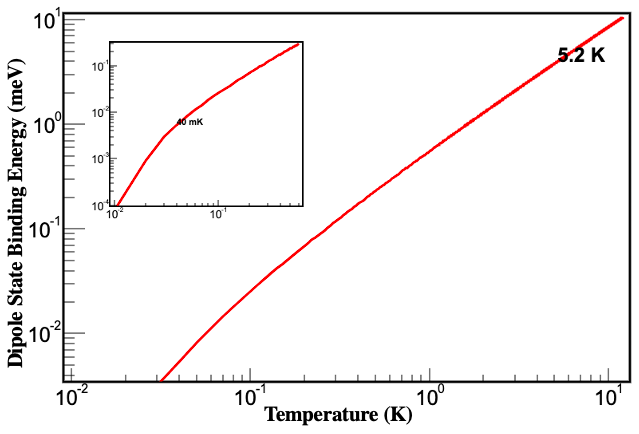}
\caption{The plot illustrates the binding energy of the dipole state, while the inset plot extends the representation across a broader temperature range. Additionally, the plot furnishes details on the binding energy of the dipole state at two distinct temperatures, namely 5.2 K and 40 mK.}
\label{dipole_bind}
\end{figure}

Utilizing Equation~\ref{bin}, we can assess the binding energy for various bound states, including dipole states and cluster dipole states, as a function of energy. As an illustrative example, Equation~\ref{bin} is employed to calculate the binding energy of dipole states. Figure~\ref{dipole_bind} portrays the binding energy of a dipole state with a dipole density of $N_a=4\times 10^{11}\text{cm}^{-3}$ across different temperatures. Two specific temperature points are highlighted in the plot to emphasize the dipole state's binding energy. At a temperature of 5.2 K, the binding energy of the dipole states is calculated at 3.98 meV. However, as the temperature drops to 40 mK, the binding energy decreases substantially to 0.005 meV, reaching a level comparable to thermal energy.

The binding energy of dipole states and cluster dipole states was measured at 5.2 K using both an n-type detector and a p-type detector in our previous study~\cite{bhattarai2023investigating, raut2023development}. The obtained values were found to be similar, as demonstrated in our research at 5.2 K~\cite{bhattarai2023investigating, raut2023development}. Generally, the binding energy of cluster dipole states is expected to differ from that of dipole states at the same temperature. This expectation arises because the binding energy of dipole states is confined by the Coulomb potential associated with the dipole size, whereas the binding energy of cluster dipole states is influenced by both the Coulomb potential determined by the trapping distance between two charge carriers with opposite charges and the dipole size. This was demonstrated by the previous measurements at 5.2 K~\cite{bhattarai2023investigating, raut2023development}, where the dipole size and the trapping distance between two mobile charge carriers with opposite charges are slightly different. However, this scenario may change when the detector is operated at 40 mK, where the dipole size becomes significantly larger than the trapping distance of two opposite charge carriers. In this scenario, the binding energy of dipole states is anticipated to be smaller than that of cluster dipole states.
To assess the binding energy of cluster dipole states at 40 mK, we utilize the data obtained from Run-67.

\subsection{Binding Energy of Cluster Dipole States }

\begin{figure}[H]
\centering
\includegraphics[clip,width=0.9\linewidth]{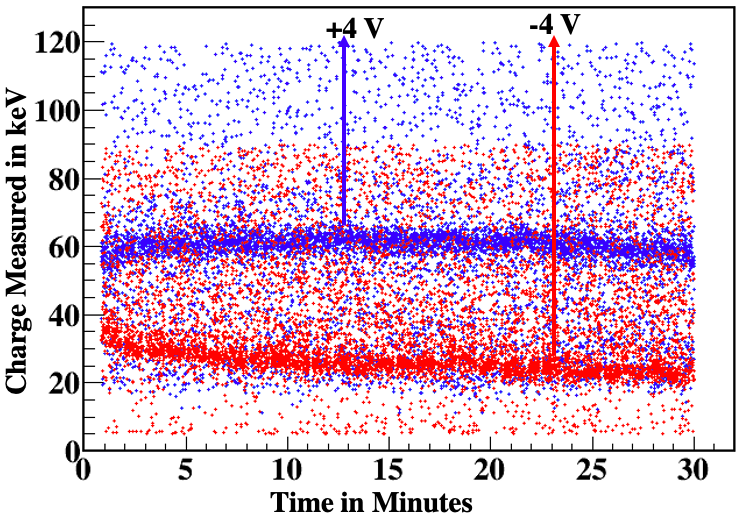}
\caption{The depicted figure showcases the dynamic charge response of Q4 over time, influenced by two opposing biases: +4 V and -4 V. These biases symbolize the migration of holes under positive bias and electrons under negative bias across the entire detector during Run-67. The data unveils that the detection of electron-hole pairs originating from Q4, induced by 59.54 keV gamma rays, is fully realized under a +4 V bias. Conversely, under a -4 V bias, the energy detection is notably lower than 59.54 keV throughout a 30-minute period. This observation suggests a substantial entrapment of electrons within the p-type Ge detector.}
\label{elect_trap}
\end{figure}

The presence of dipole states and cluster dipole states within the detector is influenced by both the detector's configuration and its operational mode. For this p-type detector, during the dilution refrigeration process, as the temperature approaches 40 mK, 100\% of impurity atoms precipitate out from the conduction band, leading to the formation of electric dipole states identified as $A^{0^{*}}$. Subsequently, electrons generated by the background radiation from the environment become trapped by the excited dipole state ($A^{0^{*}}$), resulting in the emergence of the cluster dipole state labeled as $A^{-^{*}}$. This phenomenon was observed during Run-67 when the detector was exposed to a $^{241}$Am source in two operational modes. The creation of electron-hole pairs near the surface of the detector, primarily induced by 59.54 keV gamma rays, was a key aspect. In the first mode, a negative bias was applied to the detector, prompting electrons to drift across its surface. Conversely, in the second mode, a positive bias was employed, leading to the drift of holes across the detector. For illustration, Figure~\ref{elect_trap} depicts the detector operating at both -4 V and +4 V. The figure clearly indicates that electrons become trapped, whereas holes do not exhibit the same behavior.

In the UMN K100 laboratory, we contend with the environmental background of gamma-rays inducing electron-hole pairs within the detector. Additionally, cosmic-ray muons interact with the detector, resulting in the generation of electron-hole pairs. These background electrons become ensnared in a dipole state, leading to the formation of cluster dipole states, while holes experience less entrapment in a p-type detector subjected to the same bias voltage. This discrepancy arises from the limited space available for immobilized negative ions, responsible for trapping charge carriers, compared to the ample space for mobile bound holes.

The mobility of holes is dictated by the Onsager radius, denoted as $R$ as defined earlier. Since the trapping probability is proportional to the available space where charges can be trapped, the likelihood of trapping for immobilized negative ions is smaller than that of mobile holes. Consequently, the probability of creating $A^{-^{*}}$ states is greater than that of forming $A^{+^{*}}$ states in a p-type detector. Note that $A^{+^{*}}$ and $A^{-^{*}}$ represent excited states that differ from the ground states of $A^{+}$ and $A^{-}$. This elucidates why electrons are more profoundly trapped than holes in a p-type detector, as clearly articulated in the recent publication~\cite{mei2023exploring}.

Under a sufficiently electric field, holes traverse the detector undergoing impact ionization of a dipole state or a cluster dipole state at low temperatures. At a temperature of 40 mK, despite the binding energy of dipole states being comparable to thermal energy, the impact ionization of holes takes place in conjunction with a cluster dipole state. This phenomenon arises due to Coulomb repulsion among holes, which inhibits the occurrence of this process at the operational voltage employed within the investigation, limited to a few volts.

Upon applying a positive bias voltage from the upper region of the detector and exposing it to an $^{241}$Am source positioned above, hole carriers traverse the detector. Subsequently, these drifting holes engage in impact ionization interactions with the cluster dipole states ($A^{-^{*}}$). As the bias voltage increases, the holes, serving as liberated charge carriers, accumulate greater kinetic energy, leading to the emission of electrons from the traps. This phenomenon results in a decrease in the count of cluster dipole states ($A^{-^{*}}$), inducing the transition of $A^{-}$ states, as depicted in Figure~\ref{cluster}. The amplification of a single hole to two holes per reaction ($h^{+} + A^{-^{*}} \rightarrow e^{-} + 2h^{+} + A^{-}$) tangibly demonstrates the occurrence of hole impact ionization within the detector.

Our investigation has revealed the presence of a cluster dipole state at the low cryogenic temperature of 40 mK. While the exact density of the cluster dipole state requires determination, we can utilize Equation ~\ref{bin} to calculate the binding energy associated with this state. To ascertain the density of the cluster dipole state, we will employ the relationship provided by Equation ~\ref{len}. Consequently, the updated equation for calculating the binding energy of the cluster dipole state is as follows:

\begin{equation}\label{bind_clu}
E_{BB'}=K_BT~log(\frac{N_V}{4p_{cd}}),
\end{equation}
where $E_{BB'}$ represents the binding energy for the cluster dipole state, and $p_{cd}$ denotes the density of the cluster dipole state. Considering the model in Equation ~\ref{len} to determine the density of the cluster dipole state at a cryogenic temperature of 40 mK, when the detector response is in an impact ionization mode with sufficient bias voltage and some cluster dipole states ($A^{-^{*}}$) are ionized through impact ionization, we can express Equation ~\ref{len} as $ \log(A) = p_{cd} \times \sigma_{0}E^{[p_1]}\times d$. Here, $\sigma_{imp}(E) = \sigma_0E^{[p_1]}$, with $[p_1]$ as a fitting parameter. To simplify further, let's assume $[p_0] = p_{cd} \times \sigma_0$. Now, the simplified expression for calculating the binding energy for fitting the observed data is:

\begin{equation}\label{binding}
log(A)=[p_0]\times~E^{[p_1]}\times d,
\end{equation}

\begin{figure}[H]
\centering
\includegraphics[clip,width=0.9\linewidth]{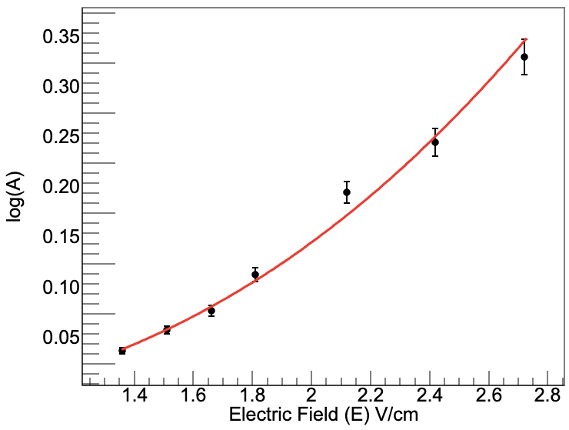}
\caption{Displayed is the logarithmic value of the amplification factor, A, obtained by summing two fitting parameters, $p_0$ and $p_1$ from Table ~\ref{t1}. $log(A)$ is plotted against the applied electric field. The error in $log(A)$ was calculated, with the error in the electric field measurement primarily influenced by the precision of the applied bias voltage.}
\label{bind}
\end{figure}

Figure ~\ref{bind} displays the fitting line of Equation ~\ref{binding}. From this, we obtained the values of two fitting parameters: $[p_0]=8.9\times10^{-3}\pm4.7\times10^{-4}$ and $[p_1]=2.5 \pm 7.7\times10^{-2}$. We know that $[p_0]$ equals $p_{cd}\times\sigma_0$. By substituting the values of $\sigma_0$ and $[p_0]$, we find $p_{cd}=3.12\times 10^{12}\times E_{BB'}~cm^{-3}$. Thus, according to Equation ~\ref{bind_clu}, the binding energy for cluster dipole states ($A^{-^{*}}$) at a cryogenic temperature can be expressed as:

\begin{equation}\label{bindc}
E_{BB'}=K_BT~log(\frac{N_V}{4\times3.12\times10^{12}~E_{BB'}}),
\end{equation}

Plugging in all the constant values and solving Equation ~\ref{bindc} numerically at a cryogenic temperature of 40 mK, we obtain $E_{BB'} = 0.034 \pm 0.0017$ meV. This ultra-low binding energy of cluster dipole states ($A^{-^{*}}$) at cryogenic temperatures suggests a delicate balance between the forces holding these clusters together. At such low temperatures, the energy states of these clusters exhibit minimal thermal agitation, revealing a fundamental characteristic of their stability.

Calculating the binding energy of the cluster dipole state enabled us to determine the density of cluster dipole states, which is found to be $p_{cd} = 3.12 \times 10^{12}~E_{BB'}$. Consequently, the density of cluster dipole states is found to be $1.05 \times 10^8\pm 5.25\times 10^{6}~cm^{-3}$. Similarly, the zero-field impact ionization cross-section for charged carriers within the detector is computed using Equation ~\ref{cross}, resulting in a value of $8.45 \times 10^{-11}\pm 4.22\times 10^{-12}~cm^2$.

\section{Conclusion}

We conducted an experimental investigation into the time-dependent impact ionization phenomenon within a p-type Ge detector. Our findings unveil that temporal fluctuations in charge amplification, particularly concerning the 59.54 keV emission line from an $^{241}$Am source, mainly arise from impact ionization induced by drifting holes traversing the detector. Our study delved into understanding the potential mechanisms driving this impact ionization, primarily attributing it to the generation and subsequent impact ionization of cluster dipole states.

Our study focused on a critical examination of binding energy and zero-field impact ionization cross-section of cluster dipole states within a low-temperature p-type Ge detector. At 40 mK, the zero-field cross-section of cluster dipole states was determined to be $8.45 \times 10^{-11}\pm 4.22\times 10^{-12}cm^2$, representing a two-order-of-magnitude increase compared to the cross-section of neutral impurity in Ge reported by Phipps et al.\cite{phipps2016observation}, which stands at $5\times 10^{-13}$$\text{cm}^2$. This discrepancy is attributed to the notable difference in binding energy between cluster dipole states at 40 mK (0.034 meV) and neutral impurity states (approximately 10 meV).
We also calculated the binding energy of cluster dipole states ($A^{-^{*}}$), resulting in $0.034 \pm 0.0017$. This ultra-low binding energy is distinctive to the detector state and plays a crucial role in the exploration of low-mass dark matter (DM) at the MeV scale. The low-scale binding energy, arising from impact ionization occurrences via intrinsic amplification, has the potential to enhance the detector's capability to achieve an ultra-low energy threshold, reaching as low as sub-meV.

The calculated cluster dipole state ($A^{-^{*}}$) binding energy, observed when the detector operated at cryogenic temperatures, is lower than the typical binding energy associated with ground-state impurities in Ge, which is $\sim$10 meV for the detector in our study. However, this binding energy is higher than the thermal energy at 40 mK, which is 0.005 meV. This suggests that these cluster dipole states might not significantly contribute to the usual impurity-related effects observed in the detector at higher temperatures. Additionally, it implies that these cluster dipole states might remain stable and distinguishable from thermal fluctuations at this low temperature. This stability hints at the potential for these states to maintain a distinct identity, unaffected by thermal agitation, thus offering a stable feature or characteristic in the detector's behavior at cryogenic temperatures.

At electric fields below 3 V/cm, our calculations yield a density of cluster dipole states ($A^{-^{*}}$), denoted as $p_{cd}$, at $1.05 \times 10^8\pm 5.25\times 10^{6}~cm^{-3}$. This value signifies the density of states associated with cluster dipole states formed by trapping electrons from background radiation. Our findings underscore the capacity of energetic charge carriers to ionize these cluster dipole states, thereby generating additional charge carriers. This phenomenon holds particular significance in facilitating impact ionization, especially in scenarios characterized by low electric fields.

In summary, our study offers crucial insights into the behavior of impurities within Ge detectors. These revelations have the potential to influence the design of innovative detectors, particularly those crafted for tasks such as MeV-scale DM searches. Additionally, the observation of low binding energies suggests the feasibility of utilizing appropriately doped impurities in Ge for the creation of low-threshold detectors capable of detecting low-mass MeV scale DM particles.

\section{Acknowledgement}
 
The authors extend their sincere gratitude to Vuk Mandic for his invaluable support in facilitating this research, providing access to his dilution refrigerator, and contributing to the data acquisition system. This research received partial support from NSF OISE 1743790, NSF PHYS 1902577, NSF PHYS 2310027, DOE DE-SC0024519, DOE grants DE-FG02-10ER46709 and DE-SC0004768, as well as a research center funded by the State of South Dakota.

\bibliographystyle{unsrt}

\bibliography{Citation}
\end{document}